\documentclass[aps,pre,superscriptaddress,twocolumn,balancelastpage,nofootinbib,10pt]{revtex4-1}

\usepackage{amsmath}
\usepackage{amssymb}
\usepackage{hyperref}
\usepackage{graphicx}
  \graphicspath{{/}{figures/}}
\usepackage{bbm}

\newcommand{\eqn} {Eq.~}
\newcommand{\eqns}{Eqs.~}

\newcommand {\avg}[1]       {\langle#1\rangle}
\newcommand {\ee}           {\mathrm{e}}
\newcommand {\ii}           {\mathrm{i}}
\newcommand {\ket}[1]       {\lvert#1\rangle}
\newcommand {\bra}[1]       {\langle#1\rvert}

\newcommand {\bramidket}[3] {\langle#1\vert#2\vert#3\rangle}
\newcommand {\abs}[1]       {\lvert#1\rvert}

\newcommand {\spinup} {\mathnormal{\uparrow}}

\newcommand {\trion}  {\mathrm{T}}

\newcommand {\half}   {\tfrac{1}{2}}

\newcommand {\Li}     {\mathcal{L}}  % Liouville operator
\newcommand {\ET}     {E_\trion}
\newcommand {\Tpulse} {T_\mathrm{pulse}}

\newcommand {\Hel}    {H_\mathrm{e}}
\newcommand {\Hnuc}   {H_\mathrm{N}}
\newcommand {\Hcoupl} {H_\mathrm{coupl}}

\newcommand  {\muB}   {\mu_\mathrm{B}}
\newcommand  {\muN}   {\mu_\mathrm{N}}
\newcommand  {\gel}   {g_\mathrm{e}}
\newcommand  {\gN}    {g_\mathrm{N}}
\newcommand  {\gNj}[1] {g_{\mathrm{N},#1}}

\newcommand  {\ps}    {\,\mathrm{ps}}
\newcommand  {\ns}    {\,\mathrm{ns}}
\newcommand  {\mus}   {\,\mu\mathrm{s}}
\newcommand  {\ms}    {\,\mathrm{ms}}

\newcommand  {\MHz}   {\,\mathrm{MHz}}
\newcommand  {\GHz}   {\,\mathrm{GHz}}
\newcommand  {\Tesla} {\,\mathrm{T}}

\newcommand {\Tr}    {\mathop{\mathrm{Tr}}\nolimits}

\begin{document}
\title{Influence of the nuclear Zeeman effect on mode locking in pulsed semiconductor quantum dots}
\date{\today}
\author{Wouter Beugeling}
\affiliation{Lehrstuhl f\"ur Theoretische Physik I, Technische Universit\"at Dortmund, Otto-Hahn-Stra\ss e 4, 44221 Dortmund, Germany}
\affiliation{Lehrstuhl f\"ur Theoretische Physik II, Technische Universit\"at Dortmund, Otto-Hahn-Stra\ss e 4, 44221 Dortmund, Germany}
\author{G\"otz S. Uhrig}
\affiliation{Lehrstuhl f\"ur Theoretische Physik I, Technische Universit\"at Dortmund, Otto-Hahn-Stra\ss e 4, 44221 Dortmund, Germany}
\author{Frithjof B. Anders}
\affiliation{Lehrstuhl f\"ur Theoretische Physik II, Technische Universit\"at Dortmund, Otto-Hahn-Stra\ss e 4, 44221 Dortmund, Germany}

\begin{abstract}
The coherence of the electron spin in a semiconductor quantum dot is strongly enhanced by mode locking through nuclear focusing, where the synchronization of the electron spin to periodic pulsing is slowly transferred to the nuclear spins of the semiconductor material, mediated by the hyperfine interaction between these. The external magnetic field that drives the Larmor oscillations of the electron spin also subjects the nuclear spins to a Zeeman-like coupling, albeit a much weaker one. For typical magnetic fields used in experiments, the energy scale of the nuclear Zeeman effect is comparable to that of the hyperfine interaction, so that it is not negligible. In this work, we analyze the influence of the nuclear Zeeman effect on mode locking quantitatively. Within a perturbative framework, we calculate the Overhauser-field distribution after a prolonged period of pulsing. We find that the nuclear Zeeman effect can exchange resonant and non-resonant frequencies. We distinguish between models with a single type and with multiple types of nuclei. For the latter case, the positions of the resonances depend on the individual $g$ factors, rather than on the average value.
\end{abstract}

\maketitle

\section{Introduction}

The electronic spins in ensembles of quantum dots in semiconductor materials, such as GaAs/InGaAs, have been proposed as possible building blocks for quantum computers \cite{LossDiVincenzo1998,ImamogluEA1999,KloeffelLoss2013}. At first glance, these systems appear to be unsuitable for this application because of the fast decoherence caused by the hyperfine coupling of the electrons to the nuclei of the constituent material \cite{Overhauser1953,MerkulovEA2002,SchliemannEA2003,DuttEA2005,BraunEA2005,HansonEA2007,UrbaszekEA2013}. However, it has been demonstrated \cite{ImamogluEA2003,HuttelEA2004,BrackerEA2005,GreilichEA2006Science,GreilichEA2007Science} that the coherence time can be vastly increased by subjecting the system to periodic optical pulses and an external magnetic field. The underlying mechanism is understood as mode locking: The spin dynamics gradually synchronizes to the pulse repetition rate \cite{GreilichEA2006Science,GreilichEA2007Science,VarwigEA2014PSS}. Non-resonant contributions eventually die out. Because the resonant frequencies are set by the pulse repetition rate only, the system becomes immune to dephasing and to small variations between individual quantum dots in the ensemble.

One can distinguish an electronic and a nuclear contribution to mode locking. The electron spin is affected directly by the pump pulses, and therefore responds rapidly: synchronization builds up after a few pulses already. The nuclei are not excited directly by the pulses, but the hyperfine interaction mediates the electronic mode locking slowly to the nuclear spins. As a result, nuclear contributions corresponding to resonant frequencies of the electronic Larmor oscillations grow, whereas non-resonant ones vanish. This phenomenon is known as \emph{nuclear focusing}, and is responsible for the long coherence times reported in experimental works \cite{GreilichEA2006Science,GreilichEA2007Science}.

The resonant Larmor oscillations are characterized by extremal electron spin polarization at the moment of each pulse. In practice, this means that roughly an integer or a half-integer number of \emph{electronic} Larmor oscillations fits into one pulse interval. Within a simplified model without nuclear Zeeman interaction, the system prefers the half-integer case \cite{BeugelingEA2016}, because the non-trivial action of the pulse is dominant over the ``idle'' pulses in the integer case. Although this model provides intuitive understanding of mode locking, the absence of the nuclear Zeeman interaction can alter the mode-locking behavior dramatically: In the presence of the nuclear Zeeman effect, the resonant frequencies may be found at the integer values \cite{PetrovYakovlev2012}, which suggests that the nuclear Zeeman coupling can introduce a $\pi$ shift that exchanges resonant and non-resonant frequencies.

In this work, we extend the perturbative method presented in Ref.~\cite{BeugelingEA2016} by including the Zeeman coupling of the nuclei to the external magnetic field. The nuclear Zeeman effect introduces frequency shifts, which we extract quantitatively: the characteristic magnetic field strength, where the \emph{nuclear} oscillations (Larmor frequency $\approx10\MHz/\mathrm{T}$) are synchronized with the pulsing ($\approx 76\MHz$), lies at a few tesla. This value is within the typical range used in pump-probe experiments \cite{GreilichEA2006PRL,GreilichEA2006Science,GreilichEA2007Science,VarwigEA2014PSS,BelykhEA2015}.

First, we consider a model where all nuclei have the same unique nuclear $g$ factor. In parallel to earlier works \cite{EconomouBarnes2014,PetrovYakovlev2012,BeugelingEA2016}, we calculate the distribution of the Overhauser field (magnetic field induced by the nuclear spins)  and observe the onset of mode locking. (Throughout this work, we shall use the term \emph{mode locking} as meaning the effect induced by \emph{nuclear focusing}, unless stated otherwise.) The peaks in this distribution, the hallmark for mode locking \cite{GreilichEA2006Science,GreilichEA2007PRB}, appear at frequencies corresponding to either an integer or a half-integer number of Larmor oscillations within one pulse period, depending on the strength of the nuclear Zeeman effect. The latter is linearly proportional to the nuclear $g$ factor as well as to the external magnetic field. We are thus motivated to study the influence of variation of these quantities.

Subsequently, we consider a model with multiple nuclear species (elements and isotopes), with different nuclear $g$ factors. In this scenario, the peak positions in the Overhauser-field distribution (OFD) depend on the individual $g$ factors, rather than on the average value. Because the $g$ factors of the Ga and As nuclei differ significantly \cite{Walchli1952,Stone2014}, it is possible that for a specific magnetic field, some nuclear species are compatible with peaks at integer, and others with half-integer resonant frequencies. We show that in this case, this competition prevents the OFD from building a well-developed peak structure.

Faraday rotation measurements in the typical pump-probe experiments resolve the time-dependent expectation values of the electron spin. The OFD cannot be measured directly, but some information can be inferred indirectly from the electron spin dynamics, more precisely its Fourier transform \cite{JaschkeEA2017preprint}. However, the latter is dominated by the electronic steady state that sets in rapidly. The effect of the nuclei (the Overhauser field) is weak, but could be extracted from the electronic dynamics by subtracting the electronic steady state, as demonstrated in the Appendix. 

This article is organized as follows. We introduce the model and the methods briefly in Sec.~\ref{secModel}. We explore the physics of the nuclear Zeeman effect in relation to mode locking in Sec.~\ref{secModeLocking}.  In Sec.~\ref{secConclusion}, we summarize our results and discuss the perspectives towards experimental verification. In the Appendix, we elaborate on the connection between the OFD and the experimentally accessible electron-spin dynamics.

\section{Model and methods}
\label{secModel}%

Our analysis is based on the central spin (Gaudin) model \cite{Gaudin1976} that governs the unitary time evolution of the central and nuclear spins. This model incorporates the coupling of the spins to the magnetic fields, as well as the hyperfine coupling between the electron spin on the one hand and each of the $N$ nuclear spins on the other hand \cite{SchliemannEA2003,KhaetskiiEA2002,*KhaetskiiEA2003}. We split the Hamiltonian 
\begin{equation}
  H = \Hel + \Hnuc + \Hcoupl
\end{equation}
into three terms,
\begin{subequations}\label{eqnHterms}
\begin{align}
  \Hel &= \hbar\lambda \hat{S}^x+\ET \ket{\trion}\bra{\trion},\label{eqnHel}\\
  \Hnuc &= \hbar\sum_{j=1}^N \Delta_j \hat{I}^x_j,\label{eqnHnuc}\\
  \Hcoupl &= \hbar\sum_{j=1}^N a_j \hat{\vec{I}}_j\cdot \hat{\vec{S}},\label{eqnHcoupl}
\end{align}
\end{subequations}
that describe the purely electronic part, purely nuclear part, and coupling, respectively. Here, $\hat{S}^\mu$ ($\mu=x,y,z$) are the spin operators for the central spin, and $\hat{I}^\mu_j$ ($j=1,\ldots,N$) are the spin operators of the $N$ nuclei.\footnote{In this work, we simplify the model by treating the nuclear spin degrees of freedom as spin-$1/2$, although in fact the Ga and As nuclei have total spin $I=3/2$.} For the electron, the coupling to the external magnetic field $\vec{B}=B\hat{x}$ is governed by the Larmor frequency $\lambda=\gel\muB B/\hbar$. The energy of the excited  trion ($\ket{\trion}$) state is $\ET$.
The nuclei couple to the magnetic field according to $\Hnuc$, where $\Delta_j=\gNj{j}\muN B/\hbar$ encodes the typical frequency for nucleus $j$. The nuclear $g$ factor $\gNj{j}$ depends on the element and the isotope. The hyperfine interaction given by $\Hcoupl$ between the central spin and nucleus $j$ has a  strength $\hbar a_j$, which is proportional to the probability density given by the electronic wave function at the position of the nucleus; here, we assume a Gaussian wave function by choosing $a_j\propto\ee^{-j/(N+1)}$ \cite{CoishLoss2004,FaribaultSchuricht2013PRL,BeugelingEA2016}.

%%%%%%%%%%%%% TABLE %%%%%%%%%%%%%
%
\begin{table}[t]
  \centering
  \begin{ruledtabular}
  \begin{tabular}{cllll}
    isotope   & $\mu/\muN$ & $I$ & $\gN$ & $\gN\muN/h$\\
              &            &     &       & [$\mathrm{MHz}/\mathrm{T}$]\\
    \hline
    $^{69}$Ga & $2.01659(5)$ & $3/2$ & $1.34439$ & $10.248$\\
    $^{71}$Ga & $2.56227(2)$ & $3/2$ & $1.70818$ & $13.021$\\
    $^{75}$As & $1.43948(7)$ & $3/2$ & $0.95965$ & $\phantom{0}7.315$\\
  \end{tabular}
  \end{ruledtabular}
  \caption{Magnetic moments $\mu$, spin quantum numbers $I$, and $g$ factors $\gN=\mu/\muN I$ of the Ga and As isotopes. The right-hand column gives the resonant frequency of the nucleus in $\mathrm{MHz}$ at $1\Tesla$. These values have been measured by nuclear magnetic resonance (NMR) experiments, and are listed in several reference tables, e.g., Refs.~\cite{Walchli1952,Stone2014}.}
  \label{tblNMR}
\end{table}
%
%%%%%%%%%%%%%%%%%%%%%%%%%%%%%%%%%

The energy and time scales of the electronic Zeeman effect are given by the \emph{effective} $g$ factor $\gel$. The actual value can vary, depending on the structure and composition of the sample \cite{BelykhEA2015}; here, we consider the typical value $\abs{\gel}=0.555$ \cite{GreilichEA2007Science}. The actual value of $\gel$ is negative, but in the following, we shall tacitly consider its magnitude only, because the sign is not relevant to our results. The value $\gel=0.555$ amounts to a Larmor frequency (per tesla of magnetic field) of $\gel\muB/h=7.77\,\GHz/\Tesla$. The nuclear Zeeman effect is much weaker due to the larger mass of the nuclei compared to the electron. Typical values of the nuclear Larmor frequencies are $\gN\muN/h\approx10\,\MHz/\Tesla$, i.e., roughly $800$ times smaller than the electronic value. For the nuclear isotopes in GaAs quantum dots, the values of $\gN$ and $\gN\muN/h$ are listed in Table~\ref{tblNMR}.

The aim of this work is to gain understanding from a model that describes the nuclear Zeeman effect in the simplest form. It should be noted that our assumption of the nuclear spin splitting $\Delta_j$ being proportional to $B$ may be violated in experiments which involve InGaAs quantum dots. In these systems, the strain-induced crystal field gives rise to an inhomogeneous quadrupole interaction that affects the splitting between the nuclear spin states significantly \cite{FlisinskiEA2010,KuznetsovaEA2014}. Since we neglect these effects in this work, comparisons between our theoretical results and experimental ones should be made with due care.

For additional simplicity, we start by considering a model with a single species of nuclei, to which we assign an effective $g$ factor of $\gN=1.2246$, which is the weighted average over $30\%$ $^{69}$Ga, $20\%$ $^{71}$Ga, and $50\%$ $^{75}$As (by number of nuclei or molar fraction) \cite{MachlanEA1986}. Then, the values of $\Delta_j$ are all equal to a single value $\Delta$, so that \eqn\eqref{eqnHnuc} simplifies to
\begin{equation}\label{eqnHnucSimplified}
 \Hnuc = \hbar\Delta\sum_{j=1}^N \hat{I}^x_j,\\
\end{equation}
The corresponding Larmor frequency per tesla is $\Delta/2\pi B = \gN\muN/h=9.337\,\MHz/\Tesla$. 

For the time evolution under periodic pulsing, we use the same method as presented in Ref.~\cite{BeugelingEA2016}. The pump pulses are applied every $13.2\ns$ and act instantaneously, as a unitary matrix operation on the central-spin Hilbert space \cite{YugovaEA2012,BarnesEconomou2011,EconomouBarnes2014}. Here, we consider $\pi$ pulses only \cite{[{Non-$\pi$ pulses are discussed in, e.g., Refs.~\cite{EconomouBarnes2014,BarnesEconomou2011}, and: }][{}]YugovaEA2009,*CarterEA2009}, and we assume that the light is circularly polarized, so only one spin species (here, $\ket\spinup$) can be excited to the trion state $\ket{\trion}$ \cite{ShabaevEA2003}. The time evolution is governed by the Lindblad equation \cite{BreuerPetruccione2007book}
\begin{subequations}\label{eqnLindblad}
\begin{equation}\label{eqnLindbladTime}
  \frac{d\rho}{dt}(t) = \Li \rho(t)
\end{equation}
with
\begin{equation}\label{eqnLiouville}
\Li \rho = -\frac{\ii}{\hbar} [H,\rho] - \gamma \left( \half b^\dagger b \rho + \half \rho b^\dagger b - b\rho b^\dagger\right)
\end{equation}
\end{subequations}
where $b=\ket{\spinup}\bra{\trion}$. The last term describes the effectively non-unitary process of the trion decay, with characteristic decay rate $\gamma\sim (400\ps)^{-1}$ \cite{GreilichEA2006PRL}.

The numerical results in this work are obtained with the perturbative method described in Ref.~\cite{BeugelingEA2016}, appropriately augmented in order to incorporate the nuclear Zeeman term, \eqn\eqref{eqnHnuc}. In this method, the basis states are chosen to be the eigenstates of $\hat{S}^x$ and $\hat{I}^x_j$, the electron and nuclear spin operators parallel to the magnetic axis ($\hat{x}$). The zeroth order of the perturbation theory is essentially the longitudinal part of the Hamiltonian, which is diagonal in the basis states, and which includes the nuclear Zeeman term $\Hnuc$ [\eqn\eqref{eqnHnuc}]. It should be stressed that the perturbation is the transverse ($y$ and $z$) part of the hyperfine action only \cite{BeugelingEA2016,KhaetskiiEA2002,*KhaetskiiEA2003,SchliemannEA2003}. At the level of the Hamiltonian, the nuclear Zeeman effect $\Hnuc$ merely induces shifts of the zeroth order eigenenergies  by
\begin{equation}\label{eqnZeemanShift1}
  z_p
   = \sum_{j=1}^N \Delta_j\bramidket{p}{I^j_x}{p}
   = \sum_{j=1}^N \Delta_j s_j^p,
\end{equation}
where $\ket{p}=\ket{s^p_1,\ldots,s^p_N}$ is the nuclear configuration, with $s^p_j=\pm\half$ being the eigenvalues of the spin operator $\hat{I}^x_j$. In the simplified case with $\Delta_j=\Delta$ for all $j$, $z_p$ can only be an integer or a half-integer multiple of $\Delta$, namely, $-(N/2)\Delta,(-N/2+1)\Delta, \ldots, (N/2)\Delta$.

In the full perturbative treatment of the Liouville operator $\Li$, the purely oscillatory contributions to the solutions of the Lindblad equation involve exponentials of the form %
$
  \exp[-\ii t(\epsilon_{p,\sigma}-\epsilon_{q,\tau})]
$,
where $\epsilon_{p,\sigma}$ are the eigenvalues of the Hamiltonian divided by $\hbar$, with $p,q$ labeling the nuclear configuration, and $\sigma,\tau$ the central-spin state. In addition, the solution has monotonically and oscillatory decaying contributions, which we may neglect here. Adding the nuclear Zeeman contribution through the substitution $\epsilon_{p,\sigma} \to \epsilon_{p,\sigma} + z_p$, we find the frequency shifts
\begin{equation}\label{eqnZeemanShift2}
  Z_{pq}
    = z_p-z_q
    = \sum_{j=1}^N \Delta_j\left(\bramidket{p}{I^x_j}{p} - \bramidket{q}{I^x_j}{q}\right)
\end{equation}
to the oscillation frequencies $\epsilon_{p,\sigma}-\epsilon_{q,\tau}$. Because the
perturbation theory is an expansion in orders of the transverse hyperfine coupling, i.e., in the number of spin flips, the $k$'th order involves shifts being $k$-fold sums of $\pm\Delta_j$. In the simplified model with one nuclear frequency $\Delta$, the shifts are exactly $k'\Delta$ with $k'=-k,\ldots,k$. We note that the nuclear Zeeman term does not only affect the frequency eigenvalues, but also the eigenvectors, which contain factors of the form $1/(\epsilon_{p,\sigma}-\epsilon_{q,\tau})$. This statement is also true for higher-order corrections to the eigenvalues. For simplicity of the argument, we will not discuss these higher orders in detail.

\section{Mode locking}
\label{secModeLocking}%
\subsection{Single nuclear species}
\label{subsecOneSpecies}%

First, we explore mode locking for a single nuclear species, where all nuclei share the same value of the $g$-factor, $\gN=1.2246$. The nuclear contribution to mode locking is conveniently studied using the distribution of the longitudinal spin operator $\hat{O}^x=\sum_j a_j \hat{I}^x_j$ \cite{EconomouBarnes2014,PetrovYakovlev2012,BeugelingEA2016}, which is proportional to the Overhauser field, the magnetic field generated by the nuclear spins, in the $x$ direction. This quantity is closely related to the electronic Larmor frequency, which equals $\lambda+O^x$ in leading order, see \eqns\eqref{eqnHel} and \eqref{eqnHcoupl}. (More details are provided in the Appendix.) To be precise, we study the histogram of values $O^x_{pp}$ in the expectation value
\begin{equation}\label{eqnExpectationValue}
  \avg{O^x}(t)=\Tr [\rho(t)\hat{O}^x] = \sum_p \rho_{pp}(t) O^x_{pp},
\end{equation}
where $\rho(t)$ is the density matrix that solves the Lindblad equation \eqn\eqref{eqnLindblad} in perturbation theory \cite{BeugelingEA2016}. The resulting histogram density at time $t$ is denoted as $\rho_t(O^x)$.

Because mode locking sets in slowly, the effect is barely larger than the discretization noise caused by the histogram binning. Thus, we do not study $\rho_t(O^x)$ directly, but instead divide out the initial distribution, and study the relative difference 
\begin{equation}\label{eqnRelativeOFD}
  \rho^\mathrm{rel}_t(O^x)=\rho_t(O^x)/\rho_0(O^x)-1.
\end{equation}
We shall refer to this quantity as the \emph{relative} OFD.

%%%%%%%%%%% FIGURE %%%%%%%%%%%
%
\begin{figure*}[tbp]
  \centering%
  \includegraphics[width=176mm]{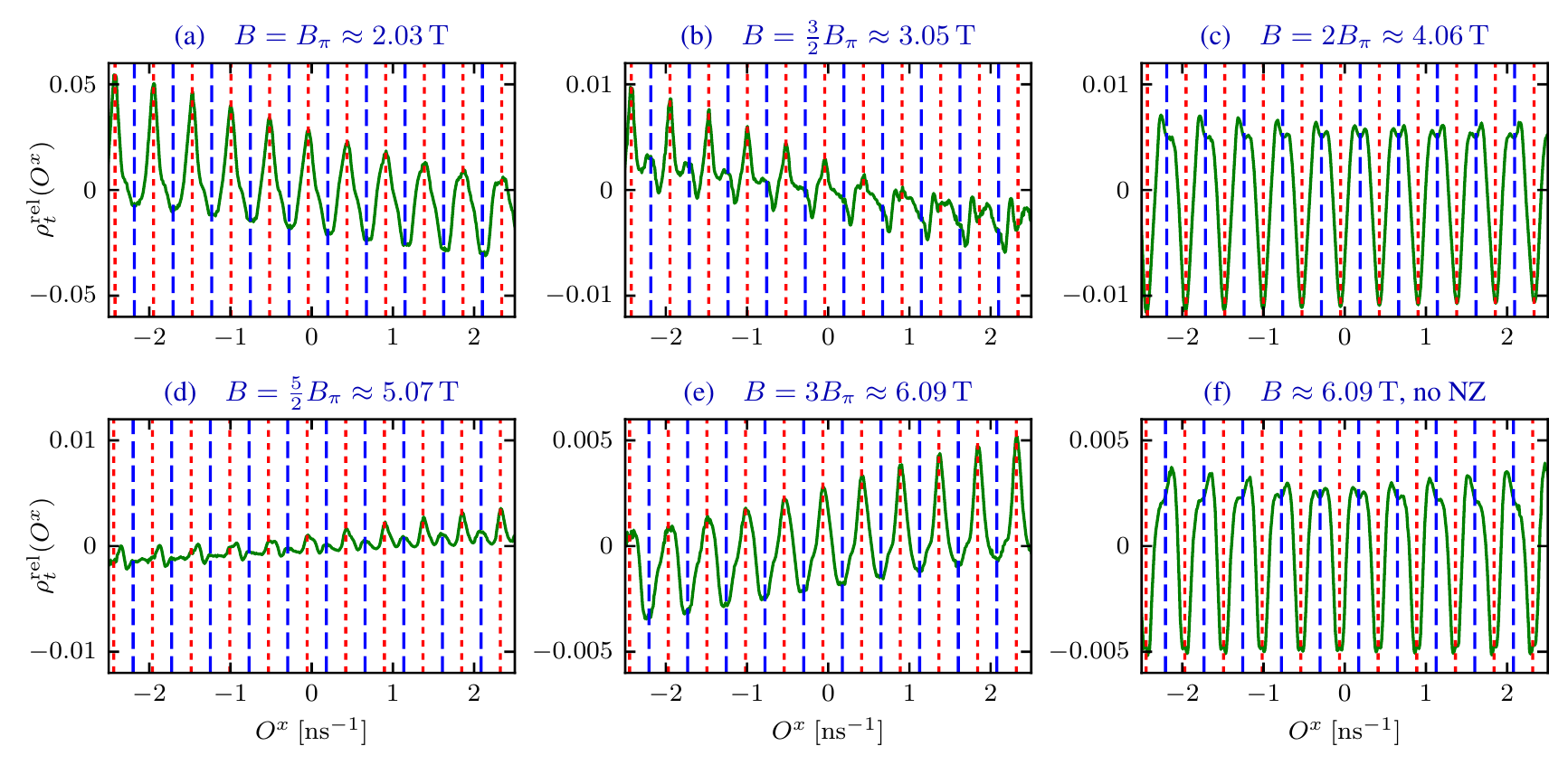}
  \caption{Relative OFDs $\rho^\mathrm{rel}_t(O^x)=\rho_t(O^x)/\rho_{0}(O^x) - 1$ with $t=1000\Tpulse$ for various values of $B$. The system size is $N=17$ and the pulse period is $\Tpulse=13.2\ns$. For (a)--(e), the $g$ factor is $\gN=1.2246$. In (f), we plot the relative OFD without nuclear Zeeman effect (no NZ) as a reference. The vertical blue (dashed) lines indicate the odd resonant frequencies, the red (dotted) lines the even ones. Note that the vertical scales differ.}
  \label{figModeLockingB}
\end{figure*}
%
%%%%%%%%%%%%%%%%%%%%%%%%%%%%%%

In Fig.~\ref{figModeLockingB}, we present the distributions of $O^x$ for several values of the magnetic field $B$. The dephasing time has been fixed at $T^*=1\ns$, and the pulsing period is $\Tpulse = 13.2\ns = 1/(75.8\MHz)$ \cite{GreilichEA2006Science}. The number of nuclei in the model is $N=17$. The resonant Larmor frequencies are given by, in leading order\footnote{The effects of the quadratic frequency shifts and the trion decay \cite{BeugelingEA2016} have been included tacitly in the calculation, but they are irrelevant for the discussion.},
\begin{equation}\label{eqnResonantCondition}
 \lambda+O^x = m\pi/\Tpulse,
\end{equation}
where even and odd values of $m$ correspond to an integer or a half-integer number of Larmor oscillations fitting between two subsequent pulses, respectively. The values of the Overhauser field $O^x$ that solve this equation are indicated by the vertical lines, blue (dashed) for odd, red (dotted) for even multiples of $\pi/\Tpulse$.

In Ref.~\cite{BeugelingEA2016}, we have demonstrated that in absence of the nuclear Zeeman effect, the OFD exhibits peaks that reside at odd values of $m$. The intuitive understanding, why odd is preferred as opposed to even, is the action of the pulse: At odd resonances, the pulse acts non-trivially by flipping the electron spin (from $\avg{S^z} < 0$ to $\avg{S^z} > 0$). At even resonant frequencies, the electron spin has performed an integer number of Larmor oscillations since the previous pulse; the pulse then acts trivially. We intuitively expect the non-trivial pulsing action (i.e., at odd resonant frequencies) to dominate. We are however unaware of a rigorous proof.

The aim of the following discussion is to investigate how the nuclear Zeeman effect changes the positions of the peaks. We draw the attention especially to the behavior at the values $B=2.03\Tesla$ and $4.06\Tesla$, see Figs.~\ref{figModeLockingB}(a) and (c). At these values, there are peaks exclusively at either even or odd multiples of $\pi/\Tpulse$, respectively. This behavior can be understood as follows. The leading order of the frequency shifts $Z_{pq}$ [\eqn\eqref{eqnZeemanShift2}] induced by the nuclear Zeeman effect is $\pm\Delta$. Also other multiples of $\Delta$ are present, but the amplitudes of these contributions are much weaker, so that they can be neglected in the perturbation theory. Thus, the ``magic'' values of the magnetic field can be obtained from equating the nuclear Larmor frequency (Zeeman energy) to the pulsing frequency
\begin{equation}\label{eqnFieldConditionEvenOdd}
 2 \Delta = n \pi / \Tpulse
\end{equation}
with integer $n$. If $n$ is odd (even), then the peaks reside at the even (odd) resonant frequencies. In particular, for $\Delta=0$, in absence of the nuclear Zeeman effect, the peaks are at the odd positions, see Fig.~\ref{figModeLockingB}(f) and Ref. \cite{BeugelingEA2016}. 

The factor of $2$ on the left-hand side of \eqn\eqref{eqnFieldConditionEvenOdd} derives from the two-spin-flip nature of mode locking: The OFD is determined by the diagonal elements of the density matrix in the spin-$x$ basis. Acting with a single spin flip onto a diagonal element yields a non-diagonal element. In order to reach a diagonal element again, an even number of spin flips is required. In Ref.~\cite{BeugelingEA2016}, this argument has been used to understand why the mode-locking \emph{rate} is quadratic in the perturbation parameters $a_j/\lambda$ (and consequently, proportional to $B^{-2}$) in leading order. This argument extends to the present case: The contribution of the nuclear Zeeman effect to the frequency associated to a matrix element of the form $\ket{p;\sigma}\bra{q;\tau}$ is (approximately) $Z_{pq}$, as stated by \eqn\eqref{eqnZeemanShift2}; a single spin flip of the nuclei thus contributes a factor $\ee^{\pm\ii t\Delta}$ in the time evolution of this matrix element. In other words, all contributions are thus shifted in frequency by $c \Delta$ with $c=-2,0,2$.\footnote{The statement that the OFD involves only \emph{even} numbers of spin flip holds in any perturbation order. In higher order, where terms with more than two spin flips play a role, $c$ may be equal to other even integers as well.} Thus, the frequency shifts of the resonances of the OFD, induced by the nuclear Zeeman effect, involve multiples of $2\Delta$ rather than of $\Delta$, which one may have expected naively based on Larmor precession of the nuclei.

For the following, we will find it convenient to denote the smallest nonzero magnetic field strength for which there are peaks only at the even resonant frequencies as $B_\pi$. Its value
\begin{equation}\label{eqnBpi}
  B_\pi = \frac{\pi \hbar}{2 \gN \muN \Tpulse} = \frac{h}{4 \gN \muN \Tpulse} \approx 2.03\Tesla.
\end{equation}
follows from solving \eqn\eqref{eqnFieldConditionEvenOdd} for $n=1$. If the external magnetic field $B$ is increased beyond $B_\pi$, the OFD alternates between resonances at odd and even frequencies with a period of $2B_\pi=4.06\Tesla$. The typical magnetic field value of $6\Tesla$ \cite{GreilichEA2007Science} approximately corresponds to $n=3$, from which even resonance frequencies are expected. The result in Fig.~\ref{figModeLockingB} is compatible with similar observations in other theoretical works \cite{PetrovYakovlev2012,JaschkeEA2017preprint}.

At intermediate fields, where \eqn\eqref{eqnFieldConditionEvenOdd} is not fulfilled for integer $n$, as in Figs.~\ref{figModeLockingB}(b) and (d), there are peaks at even \emph{and} odd multiples of $\pi/\Tpulse$. There is a continuous crossover between the even and odd cases: If one varies the magnetic field continuously from the even to the odd case, the peaks at the even resonances decrease in amplitude, approximately until halfway, i.e., where $2 \Delta \approx (n+\half) \pi / \Tpulse$. Then, peaks at the odd resonances grow, until reaching their maximum amplitude for odd integer $n$. There are no peaks at other frequencies than the even or odd resonant ones. This feature has also been reported in other studies of the nuclear Zeeman effect based on the central-spin model \cite{JaschkeEA2017preprint,Hudepohl2016_masterthesis}. 

\subsection{Mode-locking rate; dependence on $\gN$}
\label{subsecModeLockingRate}

The question arises as of whether the nuclear Zeeman term affects the rate at which mode locking sets in. We cannot answer this question from Fig.~\ref{figModeLockingB}, because the mode locking rate scales proportionally to $B^{-2}$ already in absence of the nuclear Zeeman effect \cite{BeugelingEA2016}. In order to take out the effect of the magnetic field, we fix it at $6\Tesla$, and vary the value $\Delta$ by varying $\gN$ instead. We note that this procedure is an artificial theoretical construct, which is not possible in any kind of experiment, where the $g$ factor is not a tunable variable. In theory, however, it allows us to identify the effect of the nuclear Zeeman term in a convenient manner.

\begin{figure}[tbp]
  \centering%
  \includegraphics[width=85mm]{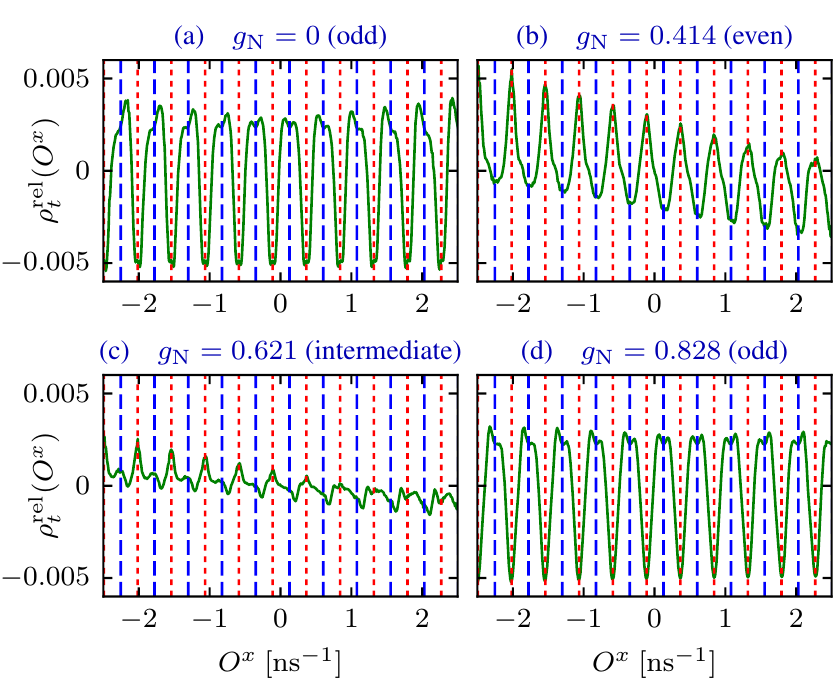}
  \caption{Relative OFDs $\rho^\mathrm{rel}_t(O^x)$ for various values of $\gN$. The magnetic field is $B=6\Tesla$ in all cases. The vertical blue (dashed) lines indicate the odd resonant frequencies, and the red (dotted) lines the even ones. The vertical scales are equal for all panels. Here, $N=17$ and $t=1000\Tpulse$.}
  \label{figModeLockingG}
\end{figure}

The results are shown in Fig.~\ref{figModeLockingG}. As we vary $\gN$, condition \eqref{eqnFieldConditionEvenOdd} is satisfied alternatingly for odd and even $n$ (even and odd resonances, respectively). The period of this alternation is $0.828$ at this magnetic field value.

%%%%%%%%%%% FIGURE %%%%%%%%%%%
%
%
%%%%%%%%%%%%%%%%%%%%%%%%%%%%%%

Comparing Figs.~\ref{figModeLockingG}(a) and \ref{figModeLockingG}(d), which satisfy \eqn\eqref{eqnFieldConditionEvenOdd} for $n=0$ and $n=2$, respectively, we observe no significant difference in the peak amplitudes. Similarly, the even case $\gN=0.414$ [$n=1$, Fig.~\ref{figModeLockingG}(b)], can be compared to the even case $n=3$ shown in Fig.~\ref{figModeLockingB}(d), at an approximately equal magnetic field. The intermediate values [Fig.~\ref{figModeLockingG}(c)] show markedly different behavior, i.e., with peaks at different positions and of different heights. Based on these observations, we conjecture that the peak structure and amplitude depends on the phase value of $2\Delta\Tpulse$ modulo $2\pi$, but not on the integer number $\lfloor 2\Delta\Tpulse/ 2\pi\rfloor$ of multiples of $2\pi$. In other words, the mode-locking rate is essentially independent of the nuclear Zeeman coupling strength $\Delta$, although the peak structure depends on the value of $2\Delta\Tpulse$ modulo $2\pi$.

In an experimental setting where the $g$ factor is fixed, but the magnetic field is varied, the mode locking rate scales roughly $\propto B^{-2}$. In presence of the nuclear Zeeman effect, the dependence is more complicated, because it is a combination of both the dependence on $\propto B^{-2}$ and the dependence on the value of $2\Delta\Tpulse$ modulo $2\pi$. 

In the long time scales typical for experiments, the mode-locking rate cannot be extracted. Instead, experiments provide information about the steady state, where the system converges to at long times. Also,  saturation effects and additional interactions beyond the present theory may play a role, e.g., the quadrupolar \cite{FlisinskiEA2010,HackmannEA2015} and dipole-dipole couplings \cite{AuerEA2009}. In contrast, the mode-locking rate is the ``speed'' at which the system converges to the steady state. Its signatures (e.g., in the amplitudes of pre-pulse and post-pulse Larmor oscillations) should be sought instead at short time scales, typically $\mus$ up to $\ms$. 

%%%%%%%%%%% FIGURE %%%%%%%%%%%
%
\begin{figure}[tbp]
  \centering%
  \includegraphics[width=86mm]{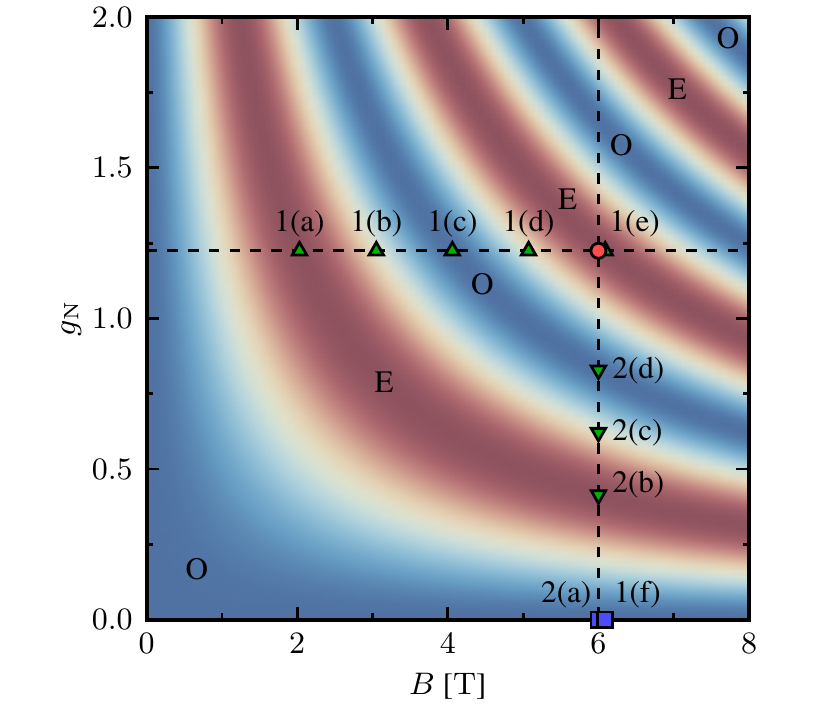}
  \caption{``Phase diagram'' for the odd (O, blue) and even (E, red) resonance conditions governed by \eqn\eqref{eqnFieldConditionEvenOdd} as function of the magnetic field $B$ and $g$ factor $\gN$. The labels odd and even correspond to the frequencies where peaks are observed in the OFD: odd and even multiples of $\pi/\Tpulse$, respectively. The color coding is determined by the values of $\cos2\Delta T_\mathrm{pulse}$ with $\Delta = g_\mathrm{N}\mu_\mathrm{N} B/\hbar$, which expresses the resonance condition \eqn\eqref{eqnFieldConditionEvenOdd} in terms of $g_\mathrm{N}$ and $B$. The horizontal dashed line corresponds to a value of $\gN=1.2246$ and the vertical dashed line to a value of $B=6\Tesla$, i.e., the ``sweeps'' that constitute Figs.~\ref{figModeLockingB} and \ref{figModeLockingG}, respectively. The panels of these figures are indicated by the green triangles with the appropriate labels. The red circle indicates a typical experimental situation at $B=6\Tesla$, and the blue squares the theoretical model without nuclear Zeeman effect, corresponding to Figs.~\ref{figModeLockingB}(f) and \ref{figModeLockingG}(a).}
  \label{figEvenOddPhaseDiagram}
\end{figure}
%
%%%%%%%%%%%%%%%%%%%%%%%%%%%%%%

For the mode locking resonance condition, only the value of $\Delta$ is relevant, not  the separate values of the magnetic field $B$ and the $g$ factor $\gN$. Varying either of those, we alternatingly enter regimes where the resonant peaks are at odd and even resonant frequencies (odd and even multiples of $\pi/\Tpulse$). In Fig.~\ref{figEvenOddPhaseDiagram}, we present a ``phase diagram'' as function of $B$ and $\gN$. The sweeps that constitute Figs.~\ref{figModeLockingB} and \ref{figModeLockingG} are represented by the horizontal and vertical dashed lines, respectively.

\subsection{Two nuclear species}
\label{subsecTwoSpecies}%
As a next step, we will lift the simplification of a single ``average'' nuclear species. Instead, we suppose the system is made up of an equal number of Ga and As nuclei. For the Ga nuclei, we take the same isotope ratio as before, i.e., $60\%$ $^{69}$Ga and $40\%$ $^{71}$Ga, which yields the average $g$ factor $\gNj{\mathrm{Ga}}=1.4899$. For As, there is only one isotope, and we read off  $\gNj{\mathrm{As}}=0.95965$ directly from Table~\ref{tblNMR}.

The $g$ factors do not only affect the couplings $\Delta_j$ of the nuclear Zeeman effect itself, but also the couplings $a_j$ between the nuclei and the central spin. For the latter, we recall that \cite{MerkulovEA2002,SchliemannEA2003}
\begin{equation}\label{eqn_aj}
  a_j = \frac{8\pi}{3}\muB\muN\gNj{j}V_0\abs{\psi(\vec{r}_j)}^2
\end{equation}
where $\abs{\psi(\vec{r}_j)}^2$ is the probability density of the central electron at nucleus $j$, $V_0$ is an appropriate volume factor \cite{MerkulovEA2002} and $\gNj{j}=\mu_j/(\muN I_j)$ is the appropriate nuclear $g$ factor. Because we are limited to small numbers of nuclei, we are interested in the correct ratio of the $a_j$'s only, and in order to keep capturing the correct collective behavior, we fix the value $\sum_j a_j^2$ such that  the dephasing time equals $T^* = 1\ns$. The distribution of the $a_j$'s is thus set up as follows. First, we distribute the values exponentially \cite{BeugelingEA2016,CoishLoss2004}, which models the Gaussian shape of the wave function $\psi(\vec{r})$. Then, the values $a_j$ ($j=1,\ldots,N$) are multiplied by $\gNj{\mathrm{Ga}}$ for odd $j$ and $\gNj{\mathrm{As}}$ for even $j$. Finally, the $a_j$ are scaled uniformly such that $\sum_j a_j^2=8/(T^*)^2$ with $T^*=1\ns$.

In this two-species scenario, the nuclei are not all resonant at the same magnetic field, i.e., for a given magnetic field, \eqn\eqref{eqnFieldConditionEvenOdd} cannot be satisfied for all nuclear species simultaneously. [We recall that the value $\gN$ implicitly appears in \eqn\eqref{eqnFieldConditionEvenOdd} as a factor in $\Delta$.] In other words, the characteristic magnetic field [cf.~\eqn\eqref{eqnBpi}] is species-specific. In this two-species model, we have $B_{\pi,\mathrm{Ga}}= h / (4 \gNj{\mathrm{Ga}} \muN \Tpulse) \approx 1.67\Tesla$ and $B_{\pi,\mathrm{As}}= h / (4 \gNj{\mathrm{As}} \muN \Tpulse) \approx 2.59\Tesla$.

%%%%%%%%%%% FIGURE %%%%%%%%%%%
%
\begin{figure}[tbp]
  \centering%
  \includegraphics[width=85mm]{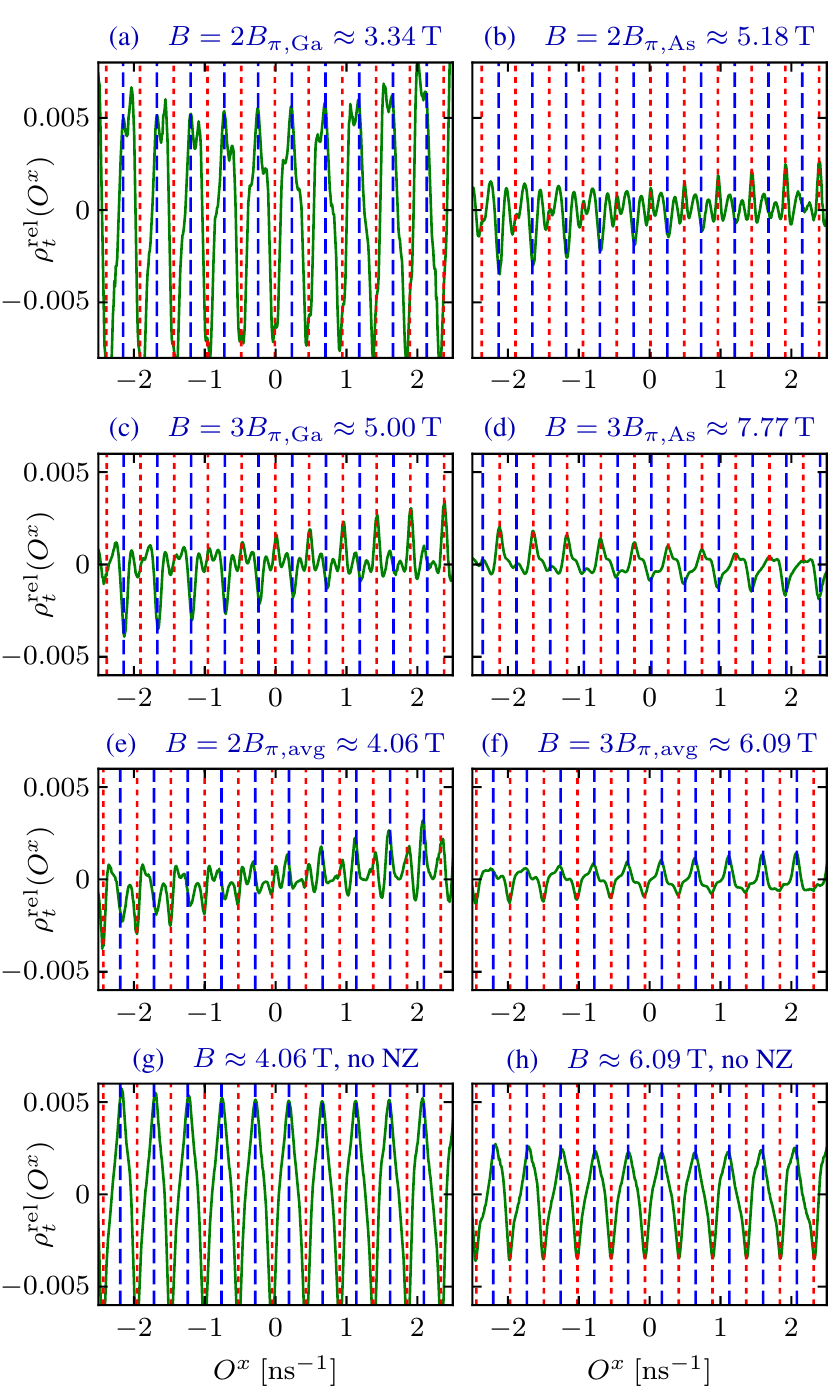}
  \caption{Relative OFDs $\rho^\mathrm{rel}_t(O^x)$ at $t=1000\Tpulse$ in the two-species model with $\gNj{\mathrm{Ga}}=1.4899$ and $\gNj{\mathrm{As}}=0.95965$, equally distributed among the $N=18$ nuclei. We probe the distribution at five different magnetic fields where some resonance condition has been fulfilled, namely,
  (a) $B=2B_{\pi,\mathrm{Ga}}\approx 3.34\Tesla$,
  (b) $B=2B_{\pi,\mathrm{As}}\approx 5.18\Tesla$,
  (c) $B=3B_{\pi,\mathrm{Ga}}\approx 5.00\Tesla$,
  (d) $B=3B_{\pi,\mathrm{As}}\approx 7.77\Tesla$,
  (e) $B=3B_{\pi,\mathrm{avg}}\approx 4.06\Tesla$, and
  (f) $B=3B_{\pi,\mathrm{avg}}\approx 6.09\Tesla$.
  Additionally, we show the OFD in absence of nuclear Zeeman coupling (no NZ) in (g) and (h), using the same magnetic-field values as in (e) and (f), respectively.
}
  \label{figModeLockingTwoSpecies}
\end{figure}
%
%%%%%%%%%%%%%%%%%%%%%%%%%%%%%%

We consider the relative difference $\rho^\mathrm{rel}_t(O^x)$ of the OFD, as before, at several magnetic-field values which correspond to some resonance condition. In Figs.~\ref{figModeLockingTwoSpecies}(a) and (b), $B=2B_{\pi,\mathrm{Ga}}$ and $B=2B_{\pi,\mathrm{As}}$, respectively, i.e., the resonance condition \eqn\eqref{eqnFieldConditionEvenOdd} is fulfilled for $n=2$. Here, we would intuitively expect peaks at the odd resonance frequencies (the blue dashed lines in the figure). By visual inspection, this prediction is certainly valid for $B=2B_{\pi,\mathrm{Ga}}$ [Fig.~\ref{figModeLockingTwoSpecies}(a)]. For $B=2B_{\pi,\mathrm{As}}=5.18\Tesla$, the peak structure is more complicated, and the strongest peaks are at the even (red dotted lines) resonance frequencies, because the odd-$n$ magnetic-field value $B=3B_{\pi,\mathrm{Ga}}=5.00\Tesla$ [see Fig.~\ref{figModeLockingTwoSpecies}(c)] lies nearby and appears to dominate. The odd-$n$ magnetic-field value $B=3B_{\pi,\mathrm{As}}=7.77\Tesla$ for As, the peaks are quite well developed. Indeed, this value of $B$ lies a considerable distance from any even-$n$ resonance (e.g., $B=2B_{\pi,\mathrm{Ga}}=6.67\Tesla$).

For reference, we include the OFD $\rho^\mathrm{rel}_t(O^x)$ for the magnetic fields $B=2B_{\pi,\mathrm{avg}}$ and $3B_{\pi,\mathrm{avg}}$, which correspond to odd and even resonances, respectively, for the average $g$ factor $\gNj{\mathrm{avg}}=1.2246$, see Figs.~\ref{figModeLockingTwoSpecies}(e) and (f). For $B=3B_{\pi,\mathrm{avg}}$ [Fig.~\ref{figModeLockingTwoSpecies}(f)], the peaks align well with the \emph{odd} resonant frequencies, although they are not so well developed as for instance in Fig.~\ref{figModeLockingTwoSpecies}(a). This is a significant difference to the one-species model, where the peaks are aligned with the \emph{even} frequencies, see Fig.~\ref{figModeLockingB}(d). For $B=2B_{\pi,\mathrm{avg}}$ [Fig.~\ref{figModeLockingTwoSpecies}(e)], neither even nor odd peaks dominate.

The OFDs in the latter two cases may be compared to the result in absence of nuclear Zeeman effect, shown in Figs.~\ref{figModeLockingTwoSpecies}(g) and (h) for additional reference. If we do not consider the nuclear Zeeman effect, the OFD is qualitatively identical to the one-species case, cf.\ Fig.~\ref{figModeLockingB}(f) versus Fig.~\ref{figModeLockingTwoSpecies}(h). The difference in the set of couplings $a_j$, determined by \eqn\eqref{eqn_aj} with either one or two values of $\gNj{j}$, does affect the OFD significantly: in both cases, the peaks are aligned with the odd resonant frequencies.

%%%%%%%%%%% FIGURE %%%%%%%%%%%
%
\begin{figure}[tbp]
  \centering%
  \includegraphics[width=85mm]{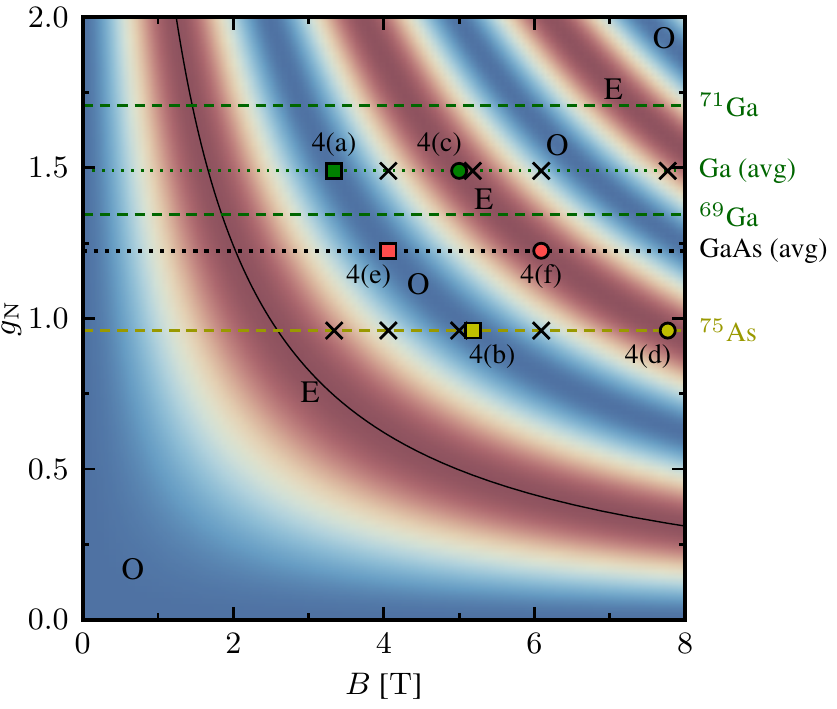}
  \caption{``Phase diagram'' for the odd (O, blue) and even (E, red) resonance conditions governed by \eqn\eqref{eqnFieldConditionEvenOdd} as function of the magnetic field $B$ and $g$ factor $\gN$, cf.\ Fig.~\ref{figEvenOddPhaseDiagram}. The horizontal dashed lines indicate the $g$ factor values of Ga and As isotopes. The dotted lines are average $g$ factors [labeled as (avg)] over the Ga isotopes and for GaAs, respectively. The solid curve expresses the relation between $\gN$ and $B_\pi$ given by \eqn\eqref{eqnBpi}. The colored markers indicate the resonances fulfilled for the cases shown in Fig.~\ref{figModeLockingTwoSpecies}(a)--(f); the crosses indicate the other isotope(s) in the two-species model for which no resonance condition is fulfilled.}
  \label{figMagnIsotopes}
\end{figure}
%
%%%%%%%%%%%%%%%%%%%%%%%%%%%%%%

In Fig.~\ref{figMagnIsotopes}, we provide a ``phase diagram'' similar to Fig.~\ref{figEvenOddPhaseDiagram}, with markers indicating the resonances and non-resonances relevant to the two-species model, with magnetic fields corresponding to the cases shown in Fig.~\ref{figModeLockingTwoSpecies}. In particular, we mention the cases $B=4.06\Tesla$ and $B=6.09\Tesla$. For $B=4.06\Tesla$, we find odd peaks in the one-species model [see red markers in Fig.~\ref{figMagnIsotopes} at $\gN=\gNj{\mathrm{avg}}$ and Fig.~\ref{figModeLockingB}(c)], 
but peaks at both even and odd frequencies in the two-species model. Indeed, the individual nuclei (Ga and As) are both on the boundary of even and odd for this value of the magnetic field, which explains the ambiguous behavior in Fig.~\ref{figModeLockingTwoSpecies}(e).
For $B=6.09\Tesla$, the OFD shows even peaks in the one-species model but odd peaks in the two-species model. We observe from Fig.~\ref{figMagnIsotopes} that both Ga and As lie in the blue area for this magnetic field, which indicates that for both species the odd resonance lies closer than the even resonance.

\subsection{Multiple species}
\label{subsecMultipleSpecies}%
The two-species results suggest that for the physics of mode locking, the specific $g$ factors are relevant. The naive simplification to a single average value of $\gN$ yields a qualitatively different OFD. Extending this idea further to a larger number of nuclear species, we find that the two-species model is also insufficient to provide reliable results, because realistically, the materials are composed of more than two isotopes. In particular, gallium contains large fractions of two isotopes ${}^{69}$Ga and ${}^{71}$Ga with significantly different nuclear $g$ factors, see Table~\ref{tblNMR}.

We have indicated the nuclear $g$ factors of the common isotopes in Fig.~\ref{figMagnIsotopes}. The nature of the resonance (peaks at even or odd frequencies) associated to each nuclear species can be read off conveniently by intersecting a constant-magnetic-field (vertical) line with the constant-$g$-factor (horizontal) line corresponding to the isotope.

For large magnetic fields ($B\gtrsim3\Tesla$), the range of $g$ factors covers multiple even/odd areas in the phase diagram, meaning that generally there will be ``even'' as well as ``odd species'' at the same field strength. With the competition between opposite types of resonances, it is difficult to predict where the peaks in the OFD will lie, or even whether there are well-developed peaks at all. This model predicts that for very small magnetic fields ($B\lesssim\frac{1}{2}B_{\pi,^{71}\mathrm{Ga}}\approx 0.72\Tesla$), the nuclear Zeeman effect is too weak for all isotopes, and thus the resonance peaks will be at odd frequencies, as predicted in the model without nuclear Zeeman effect. Interestingly, there is an intermediate region where the magnetic field $B$, approximately matches $B_\pi$ for all nuclear species, i.e., all nuclei contribute to peaks at even resonance frequencies. This region is bounded by $\frac{1}{2}B_{\pi,\mathrm{As}}\approx 1.29\Tesla$ and $\frac{3}{2}B_{\pi,^{71}\mathrm{Ga}}\approx 2.18\Tesla$. These results should be considered with due care, because the accuracy of the perturbative method is decreased in this low-field regime.

\section{Discussion and conclusion}
\label{secConclusion}%

Mode locking arises due to synchronization of the electronic Larmor oscillations (frequency $\approx \gel\muB B/\hbar$) to the pulsing frequency. The hyperfine interaction mediates this effect to the nuclei, which become ``focused'' at a sequence of resonant frequencies spaced by $2\pi/\Tpulse$. The nuclear Zeeman effect can induce a shift of the resonant frequencies. The relevant frequency scale is set by $2\Delta=2\gN\muN B/\hbar$, with the factor of $2$ deriving from the two-spin-flip nature of the mode-locking dynamics. The ratio between $2\Delta$ and the pulse frequency determines whether the mode locking peaks in the OFD are at the odd or even resonant frequencies. In addition, we find that for nuclei with different $g$ factors, the individual values are important, and that this may lead to an essentially different OFD compared to the situation where the average $g$ factor is considered. Thus, for larger magnetic fields, we cannot satisfy the resonance condition of a specific nature (odd or even) for all possible $g$ factors simultaneously. This issue is absent for smaller magnetic fields of $B\lesssim 2\Tesla$.

Unfortunately, we are unable to study the competition between odd and even in more detail, due to possible finite-size effects inherent to the method: the perturbative method is limited to small numbers of nuclei $N$, and we cannot reach values of $N$ where finite-size effects will be eliminated. Thus, we propose studies of the nuclear Zeeman effects with other methods that may reach larger values of $N$ as an interesting perspective for future research. In particular, infinite $N$ can be treated in a classical approach, which mimics the present quantum results fairly well \cite{FausewehEA2017}.

Direct measurements of the Overhauser field are elusive; the typical manner of probing the spin dynamics is through Faraday rotation and ellipticity measurements in a pump-probe configuration \cite{GreilichEA2006Science,GreilichEA2007Science,SpatzekEA2011PRB,VarwigEA2014PSS}, which typically gives access to the time evolution of the electron spin. The Fourier transform of this quantity does not correspond immediately to the OFD. The amount of mode locking \emph{in the nuclei} can be retrieved indirectly from comparison of the amplitude and phase of the electron spin Larmor oscillations before and after each pulse. In order to confirm the effects proposed here, the magnetic-field dependence of the phase shift of the electronic Larmor oscillations at the pulse must be measured. The transition from odd to even resonance conditions reported in this theoretical work should be visible as a difference of $\pi$ (half oscillation) in the phase shift. In addition, it is required that the amplitude before and after the pulse be (approximately) equal in size, in order to ensure that the mode locking in the nuclear system is sufficiently strong, and that the signatures are not mistaken for the steady-state behavior of the electron that arises on very short time scales \cite{GreilichEA2007Science,BeugelingEA2016}.

To the best of our knowledge, the predicted phase difference of $\pi$ has not been demonstrated in experiment. Measurements show that the pre-pulse phase at the pulse arrival times has a rather regular dependence on the magnetic field, with piecewise constant values across wide ranges at magnetic fields, and a sudden jump around $B=3.7\Tesla$, accompanied by a sharp reduction of the pre-pulse Larmor amplitude \cite{JaschkeEA2017preprint}.  This field strength lies within the range where our theory predicts the even-odd transitions, which suggest that the nuclear Zeeman effect may be a possible origin. However, the aforementioned requirements are not fulfilled: Firstly, the amplitude of the pre-pulse Larmor oscillations suggests that the nuclei are not strongly mode locked. Secondly, the phase values do not match the expected values $0$ or $\pi$. Finally, the magnetic field where the jump occurs appears to be independent on the pulsing frequency, which contradicts the theory exhibited in this work, cf.~\eqn\eqref{eqnBpi}. Thus, we cannot conclude that the observed phase jump originates from the nuclear Zeeman effect. Further research, both experimentally and theoretically is required in order to understand this feature. In particular, the linear dependence $\Delta=\gN\muN B/\hbar$ of the Zeeman splitting may be replaced by a more general dependence $\Delta(B)$, in order to account for nonlinearities in the splitting between nuclear spin eigenstates, caused by the nuclear quadrupolar coupling \cite{FlisinskiEA2010} and other additional interactions.

%%%%%%%%%%%%% ACKNOWLEDGEMENTS %%%%%%%%%%%%%%
\acknowledgments
We are grateful to Vasilii Belykh, Eiko Evers, Alex Greilich, and Manfred Bayer for discussing the experimental state of affairs with us.
We also thank Natalie J\"aschke for discussions about alternative theoretical approaches.
We acknowledge financial support from the Deutsche Forschungsgemeinschaft and the Russian Foundation of Basic Research in the framework of ICRC TRR 160.

%%%%%%%%%%%%% APPENDICES %%%%%%%%%%%%%%
\appendix

\section{Relation between the OFD and the electron-spin dynamics}
\label{secAppElectronSpin}
In experiments, the OFD cannot be accessed directly. Instead, mode-locking is probed using Faraday rotation and/or Faraday ellipticity measurements of the electron spin. The relation between the two is not one-to-one, but they share some common features. In this Appendix, we discuss this relation in detail, in order to provide a connection between the theoretical and experimental observations.

The basic idea of the connection between the Overhauser field $O^x$ and the electronic spin component $S^z$ is the Overhauser shift of the Larmor frequency from $\lambda=\gel\muB B/\hbar$ to approximately $\lambda+O^x$. There are additional corrections due to a phase induced by the trion decay and the transverse components $O^y$ and $O^z$ of the Overhauser field. The latter contribution is responsible for  the relation between $O^x$ and the electronic Larmor frequency being approximately, but not completely one-to-one \cite{BeugelingEA2016}.

The electron spin rapidly synchronizes to the pulsing frequency, because of its direct coupling to the pump pulses. Thus, the electron spin dynamics settles at a nearly steady state after a few ($\approx10$) pulses. From the combined action of the pulse $S^z\to -(\frac{1}{4}-\frac{1}{2}S^z)$ and of the (approximate) time evolution $S^z\to S^z \cos(\Omega\Tpulse)$, where $\Omega$ is the Larmor frequency, we find the steady-state distribution \cite{BeugelingEA2016}
\begin{align}\label{eqnSteadyStateEl}
  \bar{s}^z(\Omega) &= \frac{\cos \Omega\Tpulse}{-4+2\cos \Omega\Tpulse}\\
  \bar{s}^y(\Omega) &= \frac{\sin \Omega\Tpulse}{-4+2\cos \Omega\Tpulse}\nonumber
\end{align}
for the electronic degrees of freedom. If we consider the full system including the nuclear degrees of freedom, then the Fourier transform
\begin{equation}
 S^z(\Omega)=\int_{k\Tpulse}^{(k+1)\Tpulse} dt \ee^{-\ii \Omega t} S^z(t) 
\end{equation}
is approximately equal to $\bar{s}^z(\Omega)$ multiplied by a Gaussian envelope function from the nuclear frequency distribution, essentially the OFD $\rho_t(O^x)$.

%%%%%%%%%%% FIGURE %%%%%%%%%%%
%
\begin{figure}[tbp]
  \centering%
  \includegraphics[width=85mm]{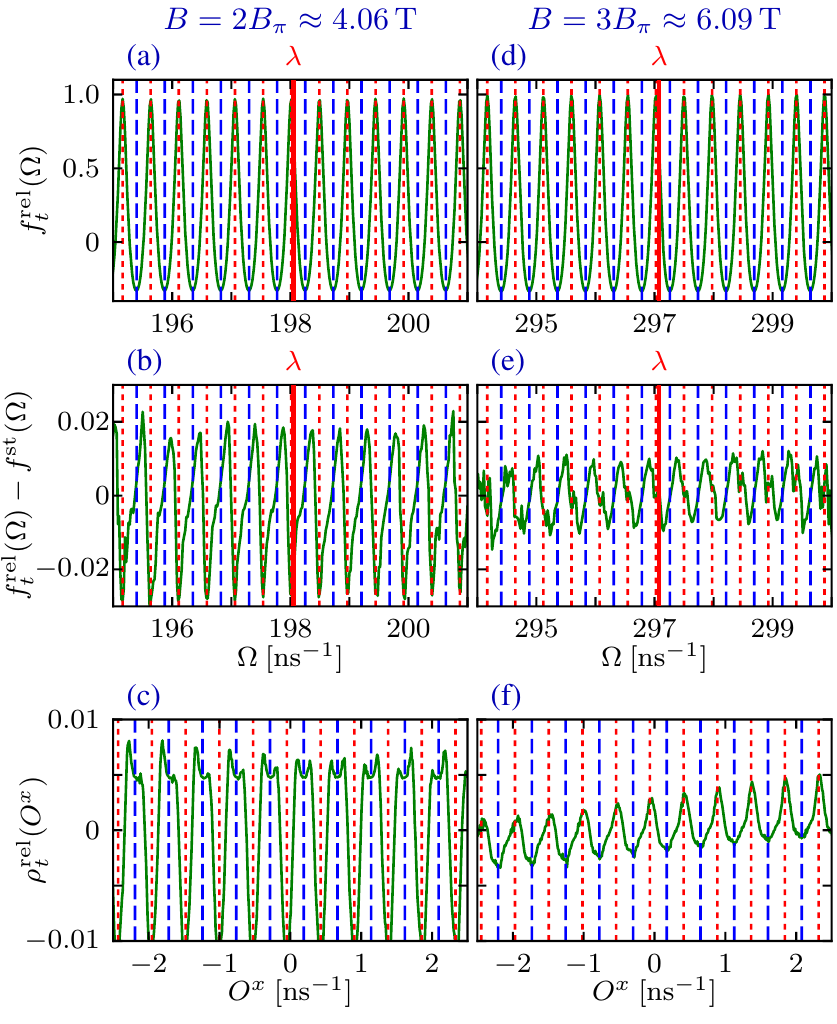}
  \caption{(a) Relative distribution $f^\mathrm{rel}_t(\Omega)$ [\eqn\eqref{eqnRelativeFT}] of the electronic spin-$z$ component [$\rho^\mathrm{rel}_t(S^z)$] as function of the frequency $\Omega$. The Larmor frequency $\lambda$ is indicated. The red (dotted) and blue (dashed) vertical lines indicate even and odd multiples of $\pi/\Tpulse$, respectively. The external magnetic field is $B=2B_\pi\approx4.06\Tesla$.  (b) The difference  between the Fourier distribution $f^\mathrm{rel}_t(\Omega)$ and the steady state $f^\mathrm{st}(\Omega)$ [see \eqn\eqref{eqnFSteady}]. (c) The corresponding relative OFD. (d)--(f) The same quantities for $B=3B_\pi\approx6.09\Tesla$. In all cases, we have $N=16$, $t=1000\Tpulse$, and $\gN=1.2246$ (single nuclear species).}
  \label{figOverhauserVsSpin}
\end{figure}
%
%%%%%%%%%%%%%%%%%%%%%%%%%%%%%%

Thus, in order to extract the effect of the nuclear mode locking from the electronic dynamics, we eliminate the dominant contributions of the Gaussian envelope and the short-term electronic steady state. 
First, we find the divide $S^z(\Omega)_t$ at large time (typically $t=1000\Tpulse$) by the initial distribution $S^z(\Omega)_0$, which represents the Gaussian envelope apart from some binning noise. Thus, we obtain the \emph{relative} Fourier distribution
\begin{equation}\label{eqnRelativeFT}
  f^\mathrm{rel}_t(\Omega) = \abs{ S^z(\Omega)_t/S^z(\Omega)_0} - 1,
\end{equation}
cf. \eqn\eqref{eqnRelativeOFD}. We discard the phase information by considering the amplitude $\abs{\bar{s}}=\sqrt{(\bar{s}^y)^2 + (\bar{s}^z)^2}$ rather than the components. For the single-species model with $N=16$ and $B=2B_\pi$, the numerically extracted relative Fourier distribution is shown in Fig.~\ref{figOverhauserVsSpin}(a). The resulting curve is almost indistinguishable from the electronic steady-state distribution
\begin{equation}\label{eqnFSteady}
 f^{st}(\Omega)
 = \abs{\bar{s}(\Omega)}/\tfrac{1}{4} -1
 = \frac{2\cos(\Omega\Tpulse)}{4-2\cos(\Omega\Tpulse)}
\end{equation}
[see \eqn\eqref{eqnSteadyStateEl}].

Next, we subtract the contribution of the electronic steady state by considering the difference $f^\mathrm{rel}_t(\Omega)-f^{\mathrm{st}}(\Omega)$, which is shown in Fig.~\ref{figOverhauserVsSpin}(b). Here, the positive values at the odd resonant frequencies and the negative values at the even ones indicate that the peaks in $f^\mathrm{rel}_t(\Omega)$ slightly decrease in amplitude compared to $f^{\mathrm{st}}(\Omega)$. The origin is the nuclear focusing; indeed, if we compare the difference $f^\mathrm{rel}_t(\Omega)-f^{\mathrm{st}}(\Omega)$ to the relative OFD [for reference, included as Fig.~\ref{figOverhauserVsSpin}(c)], we find that the peak structure is highly similar.

We also present analogous results for $B=3B_\pi$, see Figs.~\ref{figOverhauserVsSpin}(d)--(f). The relative Fourier distribution $f^\mathrm{rel}_t(\Omega)$ is again almost indistinguishable from $f^\mathrm{st}(\Omega)$ [\eqn\eqref{eqnFSteady}]. However, the difference $f^\mathrm{rel}_t(\Omega)-f^\mathrm{st}(\Omega)$ exhibits positive values at the even resonant frequencies and negative ones at the odd ones, the opposite situation from $B=2B_\pi$. This observation is compatible with the idea that the origin is nuclear, as is indeed demonstrated from the relative OFD, which has peaks at the even resonant frequencies, in this case.

Here, for relatively small degrees of mode locking, the effect on the electronic dynamics $\avg{S^z}(t)$ is small. Between Figs.~\ref{figOverhauserVsSpin}(a) and (d), the differences are unnoticeable. Peaks reside at even multiples of $\pi/\Tpulse$ in both cases, which corresponds to an even number of Larmor oscillations within the period $\Tpulse$. On the other hand, if we were able to probe the system at large times, and assume a large degree of mode locking, then in the ``odd'' case (e.g., $B=2B_\pi$) the Fourier spectrum of $f^\mathrm{rel}_t$ exhibits strong peaks at odd multiples of $\pi/\Tpulse$, and correspondingly a half-integer number of Larmor oscillations is found between two subsequent pulses. 

Thus, time-resolved measurements $\avg{S^z}(t)$ do show signatures of mode locking (nuclear focusing), but the effect is small unless the nuclei are subject to a high degree of mode locking. The degree of mode locking may be estimated from the ratio between the pre-pulse and post-pulse (negative and positive-time, respectively) amplitudes of the Larmor oscillations. Strong mode locking is characterized by nearly equal amplitudes. We stress that this ratio depends non-linearly on the size of the peaks in the OFD \cite{BeugelingEA2016}.

%\bibliography{../_csm}
%\bibliographystyle{apsrev4-1}
%\end{document}


\end{document}